\documentclass{llncs}

\usepackage{comgipp}
\usepackage{filecontents}

\usepackage[
  style=numeric,
  abbreviate=true,
  firstinits=true,
  url=false,
  isbn=false
  ]{biblatex}
\usepackage{wrapfig,booktabs} 
\usepackage[subtle]{savetrees}
\usepackage{graphicx}
\usepackage{pgfplots}
\usepackage{amsmath}
\usepackage{amsfonts}
\usepackage{amssymb}
\usepackage{stmaryrd}
\usepackage{tikz}
\usetikzlibrary{arrows,backgrounds,positioning}
\usepackage{hyperref}
\usepackage{latexml}
\usepackage{float}
\usepackage{listings}
\newcommand{\theReferenceText}{\fullcite{Cohl17}}

\usepackage{graphicx}

\hypersetup{
   colorlinks = true,
   urlcolor = blue,
   linkcolor = blue
}

\newcommand\TTt{\rule{0pt}{1.5ex}}
\newcommand\TTs{\rule{0pt}{1.8ex}}
\newcommand\TT{\rule{0pt}{2.5ex}}
\newcommand\TB{\rule[-1.2ex]{0pt}{0pt}}

\newcommand\Maple{{\sf Maple}}

\begin{document}

\title{Semantic Preserving Bijective Mappings of Mathematical 
Formulae between Document Preparation Systems and Computer Algebra Systems}

%
\author{
   Howard S. Cohl\,\inst{1}
\and
   Moritz Schubotz\,\inst{2}
\and
   Abdou Youssef\,\inst{3}
\and
   Andr\'{e} Greiner-Petter\,\inst{4}
\and
   J\"urgen Gerhard\,\inst{5}
\and
   Bonita V.~Saunders\,\inst{1}
\and
   Marjorie A.~McClain\,\inst{1}
}

\institute{\hspace{-0.15cm}Applied and Computational Mathematics Division,
NIST,
Gaithersburg, Maryland, U.S.A.,
\email{\{howard.cohl,bonita.saunders,marjorie.mcclain\}@nist.gov}
\and%
\hspace{-0.10cm}Dept.~of Computer and Information Science,
University of Konstanz, Konstanz, Germany,
\email{moritz.schubotz@uni-konstanz.de}\\
\and%
\hspace{-0.10cm}Dept.~of Computer Science, GWU,
Washington DC, U.S.A.,
\email{ayoussef@gwu.edu}\\
\and%
\hspace{-0.10cm}
DSIMG,
Technische Universit\"{a}t, Berlin, Germany,
\email{andre.greiner-petter@t-online.de}\\
\and%
\hspace{-0.10cm}Maplesoft, Waterloo, Ontario, Canada, 
\email{jgerhard@maplesoft.com}\\
\and%
\hspace{-0.10cm}Poolesville High School, Poolesville, Maryland, U.S.A.,
\email{\{joonb3,kchen1250\}@gmail.com}\\
}
\authorrunning{Cohl, Schubotz, Youssef, Greiner-Petter, Erhard, McClain, Saunders, Bang, and Chen}


\clearpage

\maketitle
\thispagestyle{firststyle}

\vspace{-0.745cm}
\begin{abstract}
Document preparation systems like \LaTeX\ offer the ability to render mathematical expressions as 
one would write these on paper.  Using \LaTeX, \LaTeXML, and tools generated for use in the National
Institute of Standards (NIST)
Digital Library of Mathematical Functions, semantically enhanced mathematical \LaTeX\ markup 
(semantic \LaTeX) is achieved by using a semantic macro set.  Computer algebra systems (CAS) 
such as {\sf Maple} and {\sf Mathematica} use alternative markup to represent mathematical 
expressions. 
By taking advantage of Youssef's Part-of-Math tagger and CAS internal representations, 
we develop algorithms to translate mathematical expressions represented in semantic \LaTeX\ to 
corresponding CAS representations and vice versa.
We have also developed tools for translating the entire {\sf Wolfram} Encoding Continued Fraction 
Knowledge and University of Antwerp Continued Fractions for Special Functions datasets,
for use in the NIST Digital Repository of Mathematical Formulae.
The overall goal of these efforts is to provide semantically enriched standard conforming 
\MathML{} representations to the public for formulae in digital mathematics libraries. 
These representations include presentation \MathML{}, content \MathML{}, generic \LaTeX, 
semantic \LaTeX, and now CAS representations as well. 

\end{abstract}

\vspace{-1.00cm}
\section{Problem and Current State}
\vspace{-0.2cm}
Scientists often use document preparation systems (DPS) to write scientific papers. The
well-known DPS \LaTeX\ has
become a de-facto standard for writing mathematics papers.
On the other hand, scientists working with formulae which occur in their research often
need to evaluate special or numerical values, create figures, diagrams and tables.
One often uses computer algebra systems (CAS), programs which provide tools for symbolic 
and numerical computation of mathematical expressions. 
DPS such as \LaTeX, try to render mathematical expressions as accurately
as possible and give the opportunity for customization of the layout of mathematical 
expressions. Alternatively, CAS represent expressions for use in symbolic computation
with secondary focus on the layout of the expressions. This difference in format is a common 
obstacle for scientific workflows.  

For example, consider the Euler-Mascheroni (Euler) constant represented by $\gamma$. Since generic 
\LaTeX{} \cite{CohlDRMF2} does not provide any semantic information, the \LaTeX\ representation 
of this mathematical constant is just the command for the Greek letter \verb|\gamma|. 
{\sf Maple} and {\sf Mathematica}, well-known CAS, represent the Euler constant $\gamma$ with 
\verb|gamma| and \verb|EulerGamma| respectively. Scientists writing scientific 
papers, who use CAS often need to be aware of representations in both DPS and CAS.  
Often different CAS have different capabilities, which implies that scientists 
might need to know several CAS representations for mathematical symbols, functions, operators, 
etc.  One also needs to be aware when CAS do not support direct translation.  
We refer to CAS translation as either the forward or backward translation respectively 
as DPS source to CAS source or vise-versa.
For instance, the CAS representation of the number $e\approx 2.71828$ (the base of the natural 
logarithm) in {\sf Mathematica} is \verb|E|, whereas in {\sf Maple} there is no directly translated symbol.  
In {\sf Maple}, one needs to evaluate the exponential function at one 
via \verb|exp(1)| to reproduce its value.

For a scientist, $\gamma$ and $e$ might represent something altogether different from these
constants, such as a variable, function, distribution, vector, etc. In these cases, it would 
need to be translated in a different way. 
In order to avoid these kinds of semantic ambiguities (as well as for other reasons), Bruce 
Miller at NIST, developing for the Digital Library of Mathematical Functions (DLMF) 
(special functions and orthogonal polynomials of 
classical analysis) project, has created a set of semantic \LaTeX{} macros 
\cite{MillerYoussef2003,miller16}.  
Extensions and `simplifications' have been provided by the Digital Repository of
Mathematical Formulae (DRMF) project.  We refer to this extended set of semantic \LaTeX\ macros 
as the DLMF/DRMF macro set, and the mathematical \LaTeX\ which uses this semantic macro set 
as semantic \LaTeX.

Existing tools which attempt to achieve CAS translations include import/export 
for \LaTeX{} expressions (such as \cite{Maple:LatexFunction,Mathematica:LatexFunction}),
as well as for \MathML.
CAS functions such as these, mostly provide only presentation translation in \LaTeX\ and do not 
provide semantic solutions or workarounds to hidden problems such as subtle differences in CAS 
function definitions. These differences may also include differences in domains or complex branch 
cuts of multivalued functions. To fill this lack of knowledge in the CAS translation 
process, one needs to provide additional information in the DPS source itself and to 
create interactive documents with references to definitions, theorems and other 
representations of mathematical expressions.  Our approach in this paper, is to develop 
independent tools for translation between different CAS and semantic \LaTeX\ representations
for mathematical expressions. 
We provide detailed information about 
CAS translation and warn about the existence 
of known differences in definitions, domains and branch cuts.
For the DRMF, we have decided to focus on CAS translation between the semantic \LaTeX\ representations 
of classical analysis and internal CAS representations for {\sf Maple} and {\sf Mathematica}.
\vspace{-0.4cm}
\subsection{A CAS, generic and semantic \LaTeX\ representation example}
\label{ACASgenericandsemantic}
\vspace{-0.2cm}
An example of a mathematical expression is $P_n^{(\alpha,\beta)}(\cos(a\Theta))$ 
where $P_n^{(\alpha,\beta)}$ is the Jacobi polynomial \cite[(18.5.7)]{NIST:DLMF}.
Table 
\ref{firsttable1} 
illustrates several DPS and CAS representations 
for this mathematical expression.
\begin{table}[t]
  \centering
\caption{DPS and CAS representations for Jacobi polynomial expression}
\vspace{0.2cm}
\begin{tabular}{ | c | c | }
\hline
\TT\TB Different Systems & Different Representations \\\hline
\TT\TB Generic \LaTeX\ & \verb|P_n^{(\alpha,\beta)}(\cos(a\Theta))| \\ \hline
\TT\TB semantic \LaTeX\ & \verb|\JacobiP{\alpha}{\beta}{n}@{\cos@{a\Theta}}| \\ \hline
\TT\TB {\sf Maple} & \verb|JacobiP(n,alpha,beta,cos(a*Theta))| \\ \hline
\TT\TB {\sf Mathematica} & \verb|JacobiP[n,\[Alpha],\[Beta],Cos[a \[CapitalTheta]]]|\\
\hline
\end{tabular}
\label{firsttable1}
\end{table}
Translating the generic \LaTeX{} representation is difficult (see \cite{CohlDRMF2}) since 
the semantic context of the $P$ is obscured.  If it represents a special function, one 
needs to ascertain which function it represents, because there are many examples of 
standard functions in classical analysis which are given by a $P$.  The semantic 
\LaTeX\ representation of this mathematical expression encapsulates the 
mostly-unambiguous semantic meaning of the mathematical expression.  This facilitates 
translation between it and CAS representations.  We use the first scan of the 
Part-of-Math (POM) tagger \cite{you17} to facilitate translation between 
semantic \LaTeX\ and CAS representations.

\vspace{-0.4cm}
\section{The Part-of-Math tagger}
\label{MLPTagger}
\vspace{-0.3cm}
There are different approaches for interpreting \LaTeX.
There exist several parsers for \LaTeX, for instance {\tt texvcjs},
which is a part of {\sf Mathoid} \cite{Schubotzmathoid}. There is also 
\LaTeXML\ \cite{LaTeXML1,MillerYoussef2003} which processes \LaTeX.
There is also an alternative grammar developed by Ginev \cite{Ginev}.
A new approach has been developed \cite{you17} which is not a fully fledged grammar but only 
extracts POM from math \LaTeX. The purpose of the POM is to extract 
semantic information from mathematics in \LaTeX. The tagger works in several stages 
(termed {\em scans}) and interacts with several machine learning (ML) based 
algorithms. 

Given an input \LaTeX{} math document, the first scan of the 
tagger examines terms and groups them into sub-expressions when indicated.
For instance \verb|\frac{1}{2}| is a sub-expression of numerator and 
denominator. A term is, in the sense of Backus-Naur form, a pre-defined 
non-terminal expression and can 
represent \LaTeX{} macros, environments, reserved symbols 
(such as the \LaTeX\ line break command \verb|\\|) or numerical or alphanumerical 
expressions. Sub-expressions and terms get tagged due the first scan of 
the tagger, with two separate tag categories: (1) definite tags (such as 
{\em operation}, {\em function}, {\em exponent}, etc.) that the tagger 
is certain of; and tags which consist of alternative and tentative features
which include alternative roles and meanings. These second category of tags
are drawn from a specific knowledge base which has been collected for the 
tagger.  Tagged terms are called math terms. Math terms are rarely distinct at 
this stage and often have multiple features.

Scans 2 and 3 are expected to be completed in the next 2 years.  These involve some 
natural language processing (NLP) algorithms as well as ML-based algorithms
\cite{Schubotzetal16,
DBLP}.
Those scans will:~(1) select the right features from 
among the alternative features identified in the first scan; (2) disambiguate 
the terms; and (3) group subsequences of terms into unambiguous sub-expressions 
and tag them, thus deriving definite mostly-unambiguous semantics of math terms 
and expressions. The NLP/ML 
algorithms include math topic modeling, math context modeling, math document 
classification (into various standard areas of math), and definition-harvesting 
algorithms.

Specifically, to narrow down the role/meaning of a math term, it helps to know which
area of mathematics the input document is in. This calls for a {\em math-document classifier}.
Furthermore, knowing the topic, which is more specific than the area of the document, will
shed even more light on the math terms. Even more targeted is the notion of {\em context} which,
if properly formulated, will take the POM tagger a long way in narrowing down the tag choices.

In \cite{you17}, Youssef defines a new notion of a math-term's context, which involves several
components, such as (1) the area and topic of the term's document; (2) the document-provided 
definitions; (3) the topic model and theme class of the term's {\em neighborhood} in the 
document; (4) the actual mathematical expression containing the term; as well as (5) a small number 
of natural language sentences surrounding the mathematical expression.
Parts of this context are the textual definitions and explanations
of terms and notations which can be present or absent from the input document. These can also 
be near the target terms or far and distributed from them. The NLP/ML-based algorithms for the 
2\textsuperscript{nd} and 3\textsuperscript{rd} scans of the tagger will model and track the 
term's contexts, and will harvest 
definitions and explanations and associate them with the target terms.

\vspace{-0.5cm}
\section{Semantic \LaTeX{} to CAS translation}\label{sec:LatexToCAS}
\begin{figure}[t]
\centering
\includegraphics[clip, trim=0.2cm 0.2cm 0.2cm 0.2cm, width=0.95\textwidth]{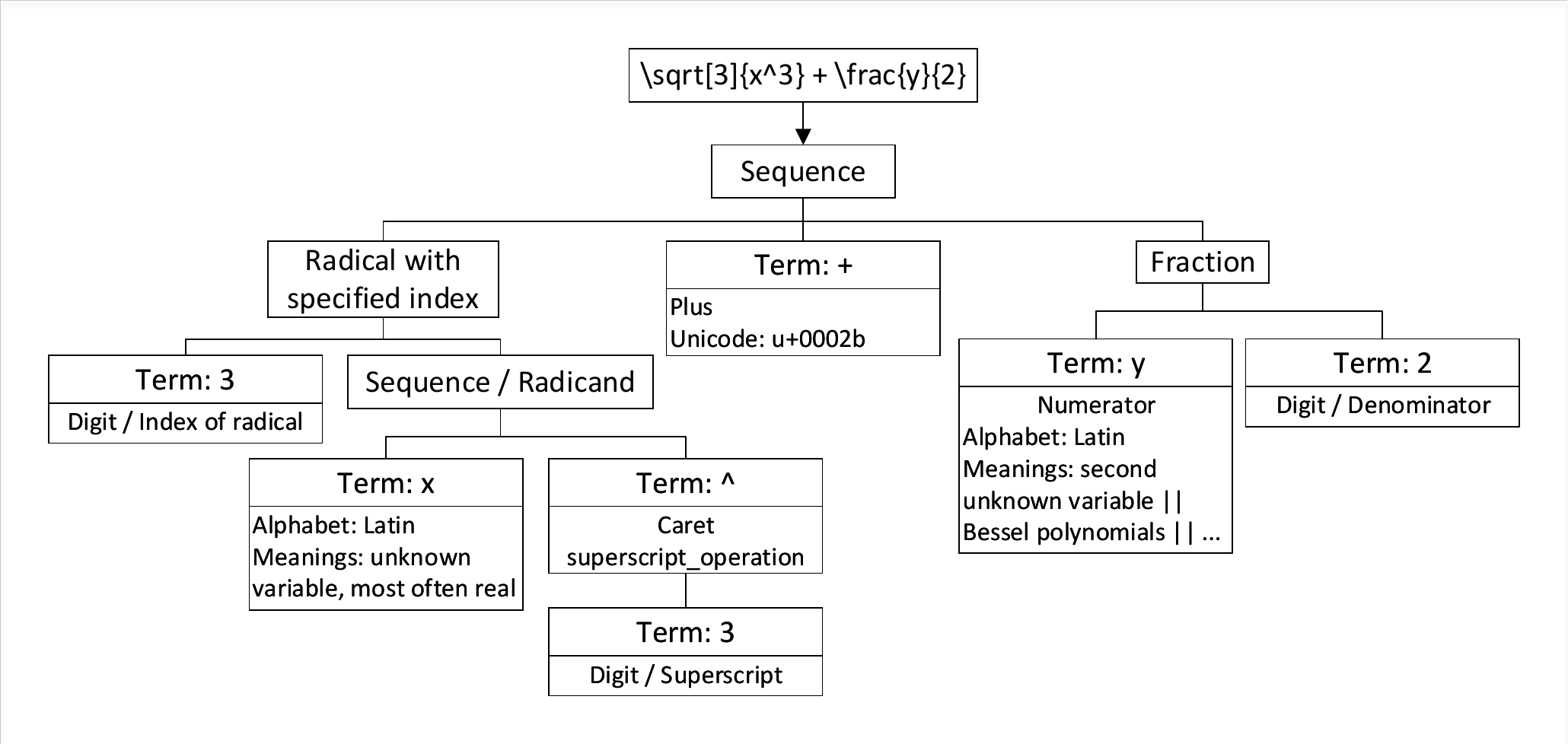}
\caption{Syntax tree of 
$\sqrt[3]{x^3} + \frac{y}{2}$
produced by the first scan of the POM tagger.}
\label{syntaxtree}
\end{figure}

\vspace{-0.3cm}
We have used a mathematical language parser (MLP) as 
an interface for the above-described first scan of the POM tagger 
to build syntax trees of mathematical expressions in \LaTeX{} and provide CAS translations from 
semantic \LaTeX{} to CAS representations.
The MLP provides all functionality to interact with the results of the POM
tagger.
We extended the general information of each term to its CAS representation, 
links to definitions on the DLMF/DRMF websites, as well as the corresponding 
CAS websites. We also add information about domains, position of branch cuts and 
further explanations if necessary. Since the multiple scans of the POM tagger 
are still a work in progress, our CAS translation is based on the first scan (see 
\S \ref{MLPTagger}). Fig.~\ref{syntaxtree} shows the syntax tree corresponding to
the \LaTeX\ expression \verb|\sqrt[3]{x^3} + \frac{y}{2}|; note that `\verb|x|' 
and `\verb|3|' in `\verb|x^3|' are not treated (in Fig.~\ref{syntaxtree}) as siblings 
(i.e., children of `\verb|^|') because the first scan of the tagger does not recognize 
this hierarchy (but it will be rectified in POM Scans 2 and 3). 
The general CAS translation process translates 
each node without changing the hierarchy of the tree recursively.
With this approach, we are able to translate nested function calls.

\begin{figure}[t]
\centering
\includegraphics[clip, trim=0.2cm 0.2cm 0.2cm 0.2cm, width=0.95\textwidth]{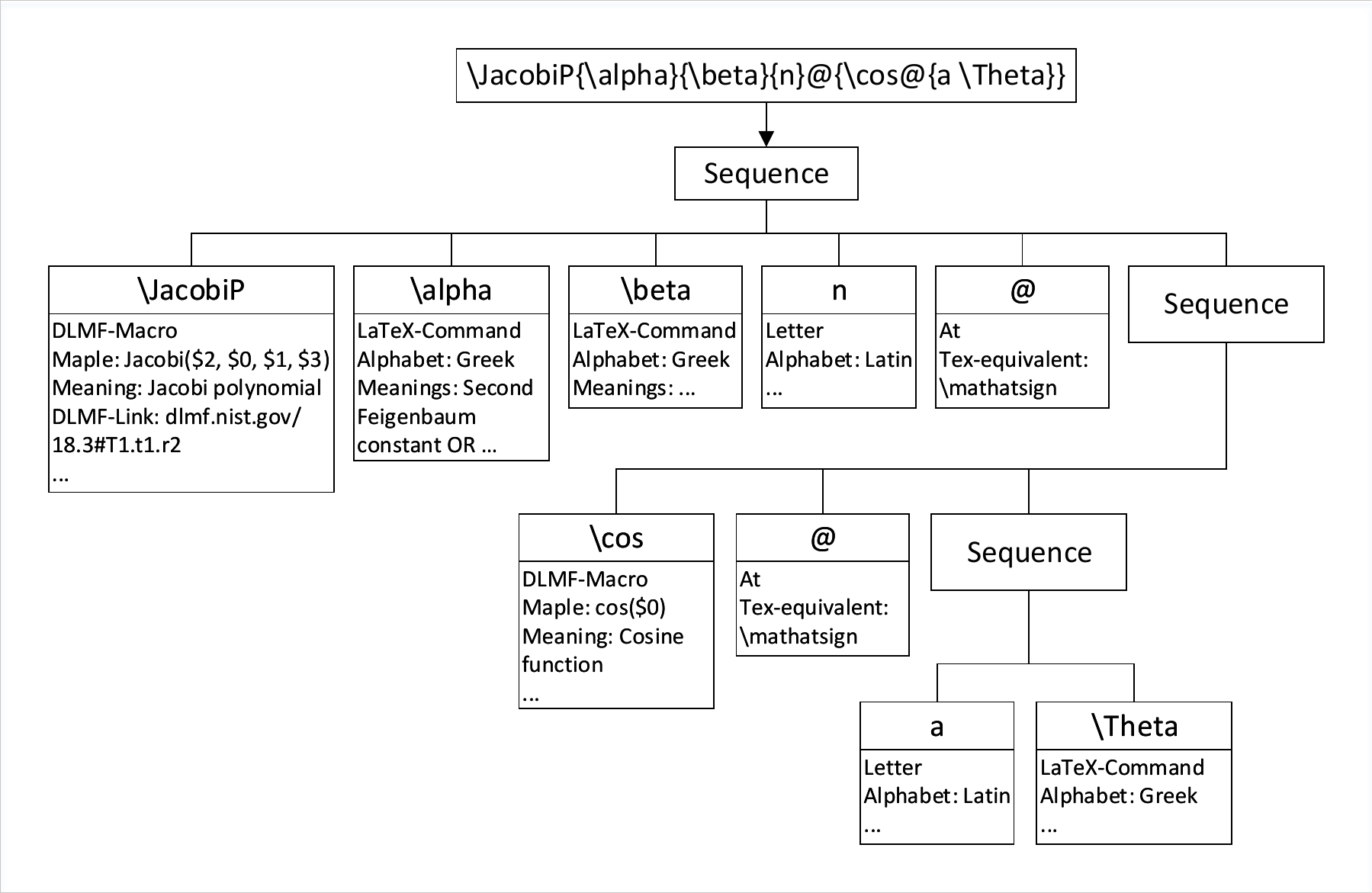}
\caption{Syntax tree for Jacobi polynomial expression
generated by the first POM scan.}
\label{syntaxtreeUseCase}
\end{figure}

The syntax tree obtained by the first POM scan depends on the known terms 
of the tagger. Although the tagger's first scan tags macros if those macros' 
definition are provided to it, it is currently agnostic of the DLMF/DRMF macros. 
Therefore, as it currently stands, the first scan of the tagger extracts, but does not 
recognize/tag DLMF/DRMF macros as hierarchical structures, but rather treats those 
macros as sequences of terms.  The syntax tree in Fig.~\ref{syntaxtreeUseCase} was created 
by the tagger for our Jacobi polynomial example in \S\ref{ACASgenericandsemantic}.
The tagger extracts expressions enclosed between open and closed curly 
braces\phantom{,}\{...\} which we refer to as {\it delimited balanced expressions}. 
The given argument is a sub-expression 
and produces another hierarchical tree structure.

\vspace{-0.4cm}
\subsection{Implementation}
\vspace{-0.1cm}
CAS translations for DLMF/DRMF macros are stored in {\tt CSV} files, to make them easy 
to edit. Besides that, CAS translations for Greek letters and mathematical 
constants are stored separately in JSON files. In addition to the DLMF/DRMF macro 
set, generic \LaTeX{} also provides built-in commands for 
mathematical functions, such as \verb|\frac| or \verb|\sqrt|.  CAS 
translations for these macros are defined in another JSON file. 

Since the POM tagger assumes the existence of special formatted lexicon files to
extract information for unknown commands, the {\tt CSV} files containing 
CAS translation information has to be converted into lexicon files. Table \ref{sineLexiconEntry}
shows a part of the lexicon entry for the DLMF/DRMF macro 
\verb|\sin@@{z}|\footnote{The usage of multiple @ symbols in Miller's \LaTeX\ macro set provides capability
for alternative presentations, such as $\sin(z)$ and $\sin\;z$ for one and two @ symbols respectively.}.
Translations to CAS are realized by patterns with placeholders. The symbol \verb|$i|
indicates the $i$-th variable or parameter of the macro.

\begin{wraptable}[8]{r}[-.5cm]{5.0cm}
\vspace{-0.8cm}
\caption{A lexicon entry.}
\begin{tabular}{| l | l |}
\hline
\TT\TB DLMF & \verb|\sin@@{z}| \\\hline
\TT\TB DLMF-Link & \href{http://dlmf.nist.gov/4.14\#E1}
{dlmf.nist.gov/4.14\#E1}\\\hline
\TT\TB \Maple{} & \verb|sin($0)|\\\hline
\TT\TB {\sf Mathematica} & \verb|Sin[$0]|\\\hline
\end{tabular}
\label{sineLexiconEntry}
\end{wraptable}

Our CAS translation process is structured recursively. A CAS translation
of a node will be delegated to a specialized class for certain kinds of nodes.
Even though our CAS translation process assumes semantic \LaTeX{} with DLMF/DRMF macros,
we sometimes allow for extra information obtained from generic \LaTeX{} expressions.
For instance, we distinguish between the following cases: (1) a Latin letter is used 
for an elementary constant; (2) a generic \LaTeX{} command (such as the \LaTeX\ 
command for a Greek letter) is used for an elementary constant. 
In both cases, the program checks if there are known 
DLMF/DRMF macros to represent the constant in semantic \LaTeX{}.
If so, we inform the user of the DLMF/DRMF macro for the constant,
but the Latin letter or \LaTeX\ command is not translated.  

There are currently only three known Latin letters where this occurs, 
the imaginary unit $i$, 
Euler's number $e$,
and Catalan's constant $C$.
If one wants to translate the Latin letter to the constant, 
then one needs to use the designated macro. In these three 
cases they are \verb|\iunit|, \verb|\expe| and \verb|\CatalansConstant|.
Examples of \LaTeX\ commands which may represent elementary constants are
$\pi$ and $\alpha$ which are often used to represent the ratio of a circle's
circumference to its diameter, and the fine-structure constant respectively
which are \verb|\cpi| and \verb|\finestructure|.
Hence, Latin and Greek letters will be always translated as Latin and Greek letters
respectively.

\begin{figure}[t]
\centering
\includegraphics[clip, trim=0.2cm 0.2cm 0.2cm 0.2cm, width=0.9\textwidth]{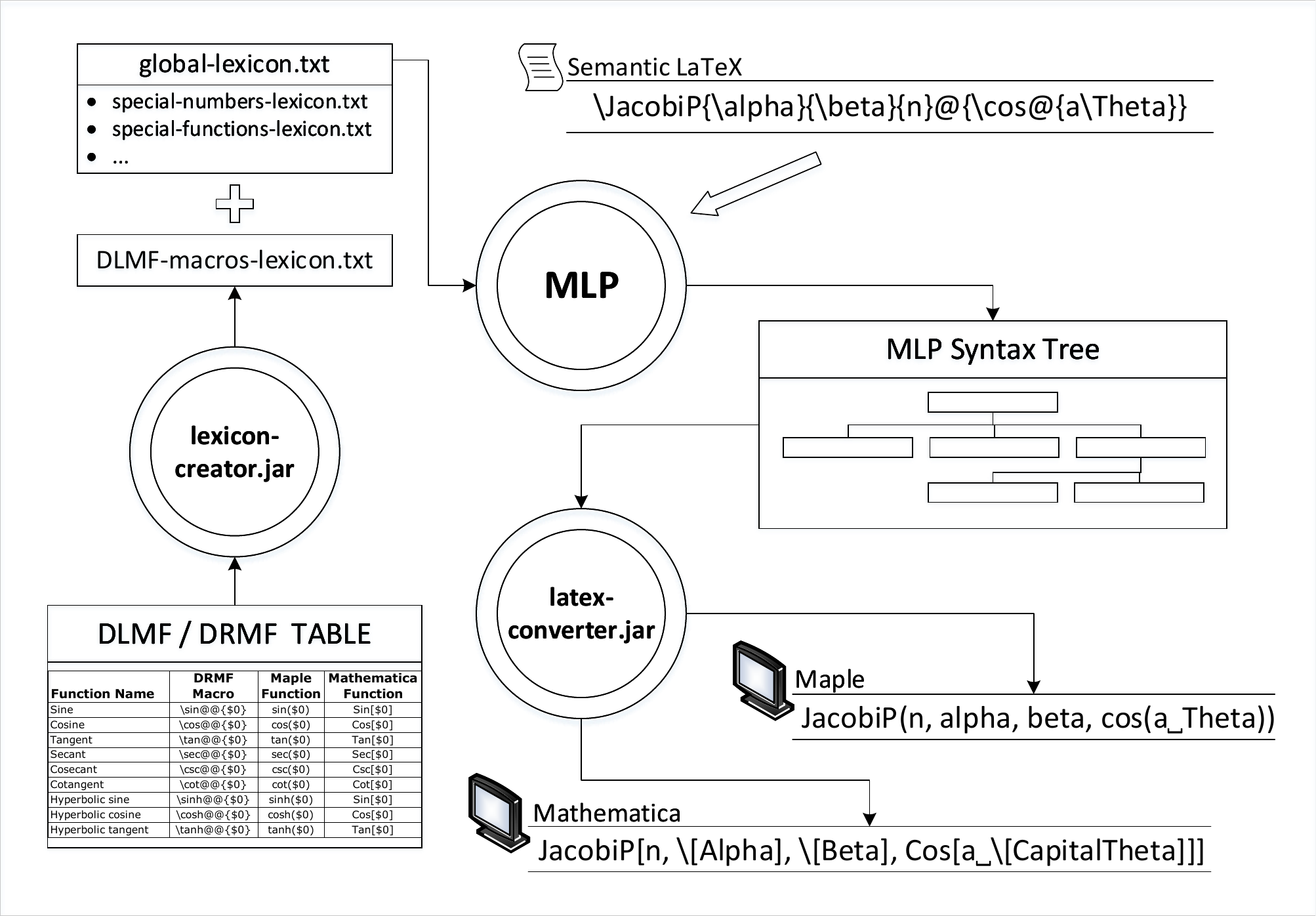}
\caption{Flow diagram for translation between 
semantic \LaTeX\ and a CAS representations. The MLP is the only
interface to the POM tagger and provides all functionality for
interaction with the results of the POM tagger (such as analyzing the
syntax tree and extracting information from the lexicon.)}
\label{translationflowchart}
\end{figure}

The program consists of two executable JAR files. One organizes 
the transformation from {\tt CSV} files to lexicon files, while the other 
translates the generated syntax tree to a CAS representation. 
Fig.~\ref{translationflowchart} describes the CAS translation process. The program 
currently supports forward CAS translations for \Maple{} and {\sf Mathematica}.

\section{{\sf Maple} to semantic \LaTeX{} translation}
\Maple{} has its own syntax and programming language, and users
interact with \Maple{} by entering commands and expressions in
\Maple{} syntax. For example, the mathematical expression
${\displaystyle \int_0^\infty (\pi+\sin(2x))/x^2 dx,}$
would be entered in \Maple{} as \\[-0.25cm]
\begin{equation}
  \label{eq:2nd}
  \texttt{int((Pi+sin(2*x))/x\^{}2, x=0..infinity)}.
\end{equation}
In the sequel, we will refer to \Maple{} syntax such as the syntactically 
correct format (\ref{eq:2nd}) as (i) the {\tt 1D} {\sf Maple} representation. 
\Maple{} also provides a (ii) {\tt 2D} representation (whose internal format
is similar to \MathML{}), and its display is similar to the \LaTeX{} rendering of 
the mathematical expression. In addition, \Maple{} uses two internal
representations (iii) \verb|Maple_DAG|,
and (iv) {\verb|Inert_Form|} representation. 
Note that, even though DAG commonly refers to the general graph theoretic/generic 
data structure, {\it directed acyclic graph}, in {\sf Maple} it has become synonymous 
with ``{\sf Maple} internal data structure,'' whether it actually represents a DAG 
or not.

\begin{wrapfigure}[19]{r}[-0.4cm]{5.5cm}
\vspace{-0.2cm}
\begin{center}
\includegraphics[clip, trim=0.5cm 0.51cm 0.5cm 0.51cm, width=0.48\textwidth]{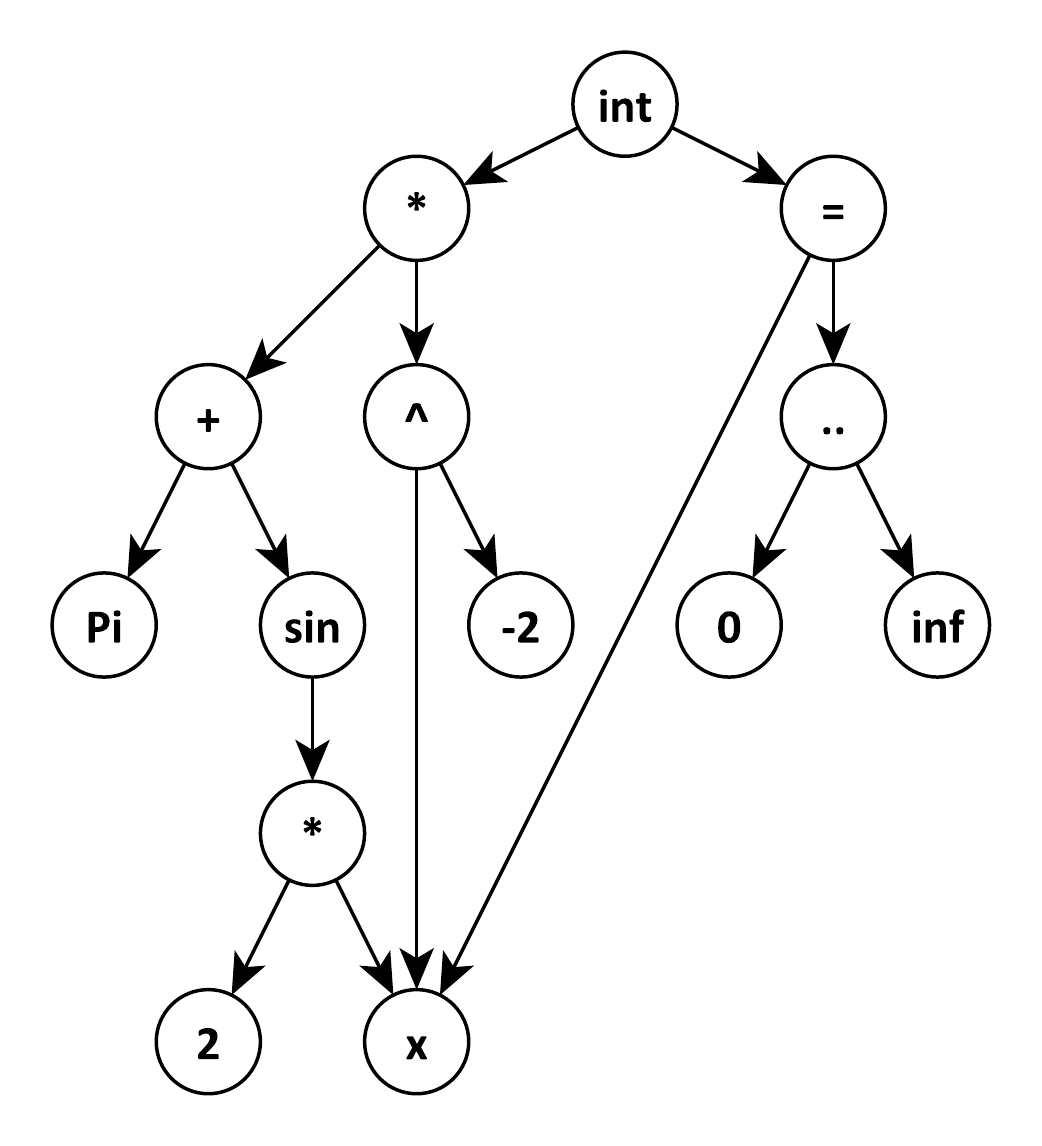}
\caption{Example {\tt Maple\textunderscore DAG} for (\ref{eq:2nd}).}
\label{DAGfigit}
\end{center}
\vspace{-0.4cm}
\end{wrapfigure}

In our translation from \Maple{} to semantic \LaTeX{}, only the \Maple{} {\tt 1D} 
and \verb|Inert_Fo|-\\
\verb|rm| representations are used.
Programmatic access to the \Maple{} kernel (its internal data structures/commands) 
from other programming
languages such as {\sf Java} or {\sf C} is possible through a published
application programming interface (API) called {\sf OpenMaple} \cite[\S 14.3]{MPG}.
The {\sf OpenMaple} {\sf Java} API is used in this project.  Some of
the functionality used includes
(1) parsing a string in {\tt 1D} representation and converting
it to its \verb|Maple_DAG| and {\verb|Inert_Form|} representations (see below);
(2) accessing elements of \Maple's internal data structures;
(3) performing manipulations on \Maple{} data structures
  in the \Maple{} kernel.

Mathematical expressions in \Maple{} are internally represented as
\verb|Maple_DAG| representations. Fig.~\ref{DAGfigit} illustrates
the {\verb|Maple_DAG|} representation of the 1D \Maple{} expression (\ref{eq:2nd}). 
The variable $x$ is stored only once in memory, and all three occurrences of it
refer to the same \Maple{} object. This type of common subexpression
reuse is the reason why \Maple{} data structures are organized as {\verb|DAG|}s
and not as trees. 
In addition to mathematical expressions, \Maple{} also has a variety
of other data structures (e.g., sets, lists, arrays, vectors,
matrices, tables, procedures, modules). The structure of a \verb|Maple_DAG| is in the form
\setlength\unitlength{1cm}
\hspace{-1mm}\raisebox{-1mm}
{\framebox(1.30,0.45){{\it Header}\hspace{2pt}}\hspace{0.1mm}%
\framebox(1.1,0.45){{\it Data$_1$}\TTs}\hspace{-1.0mm}
\framebox(0.4,0.45){$\cdots$\TTt}\hspace{-1.0mm}
\framebox(1.1,0.45){{\it Data$_n$\TTs}}}\,.\hspace{0.035cm}
{\it Header} encodes both the type and the length $n$ of the {\verb|Maple_DAG|} and
$Data_1,$\hspace{0.05cm}$\dots,$\hspace{0.05cm}$Data_n$ are {\verb|Maple_DAG|}s (see \cite[Appendix A.3]{MPG}). 

For this project, another tree-like representation that closely
mirrors the internal {\verb|Maple_DAG|} representation (and can be accessed more
easily through the {\sf OpenMaple} {\sf Java} API) was chosen, the 
{\verb|Inert_Form|}. The {\verb|Inert_Form|} is given by nested function calls 
of the form \verb|_Inert_XXX|$\,\,(Data_1,\ ...,\ Data_n)$,
where \texttt{XXX} is a type tag (see \cite[Appendix A.3]{MPG}), and
$Data_1,\ \dots,\ Data_n$ can themselves be {\verb|Inert_Form|}s. In \Maple, the
{\verb|Inert_Form|} representation can be obtained via the command
\texttt{ToInert}. For example, the {\verb|Inert_Form|} representation of the
\Maple{} expression (\ref{eq:2nd}) is
\begin{sloppypar}
\hspace{-0.3cm}\noindent {\scriptsize \texttt{\_Inert\_FUNCTION( \_Inert\_NAME("Int"), \_Inert\_EXPSEQ( 
\_Inert\_PROD( \_Inert\_SUM( \_Inert\_NAME("Pi"), }}
\end{sloppypar}
\begin{sloppypar}
\hspace{-0.3cm}\noindent {\scriptsize \texttt{\_Inert\_FUNCTION( \_Inert\_NAME("sin"), \_Inert\_EXPSEQ( \_Inert\_PROD( 
\_Inert\_NAME("x"), }}
\end{sloppypar}
\begin{sloppypar}
\hspace{-0.3cm}\noindent {\scriptsize \texttt{\_Inert\_INTPOS(2))))), \_Inert\_POWER( \_Inert\_NAME("x"), \_Inert\_INTNEG(2))) }}
\end{sloppypar}
\begin{sloppypar}
\hspace{-0.3cm}\noindent {\scriptsize \texttt{\_Inert\_EQUATION( \_Inert\_NAME("x"), \_Inert\_RANGE( \_Inert\_INTPOS(0), 
\_Inert\_NAME("infinity")))))}}.
\end{sloppypar}
\noindent In order to facilitate access to the {\verb|Inert_Form|} from the {\sf OpenMaple} {\sf Java} API,
the {\verb|Inert_Form|} is converted to a {\tt nested list} representation, where the first element 
of each (sub)-list is an \texttt{\_Inert\_XXX} tag. For example, the \Maple{} equation
\texttt{x=0..infinity} which contains the integration bounds (which is a sub-{\verb|Maple_DAG|}
of \Maple{} expression (\ref{eq:2nd})), is as follows in the {\tt nested list} representation
of the {\verb|Inert_Form|}:\\
\noindent{\scriptsize\texttt{\phantom{\hspace{-0.27cm}|}
[\_Inert\_EQUATION, [\_Inert\_NAME, "x"], [\_Inert\_RANGE, [\_Inert\_INTPOS, 0], [\_Inert\_NAME, "infinity"]]]\hspace{-0.015cm}.
}}

\subsection{Implementation}\label{sec:backwardTranslationImplementation}
\begin{figure}[b]
\centering
\includegraphics[clip, trim=0.2cm 0.2cm 0.2cm 0.0cm, width=0.8\textwidth]{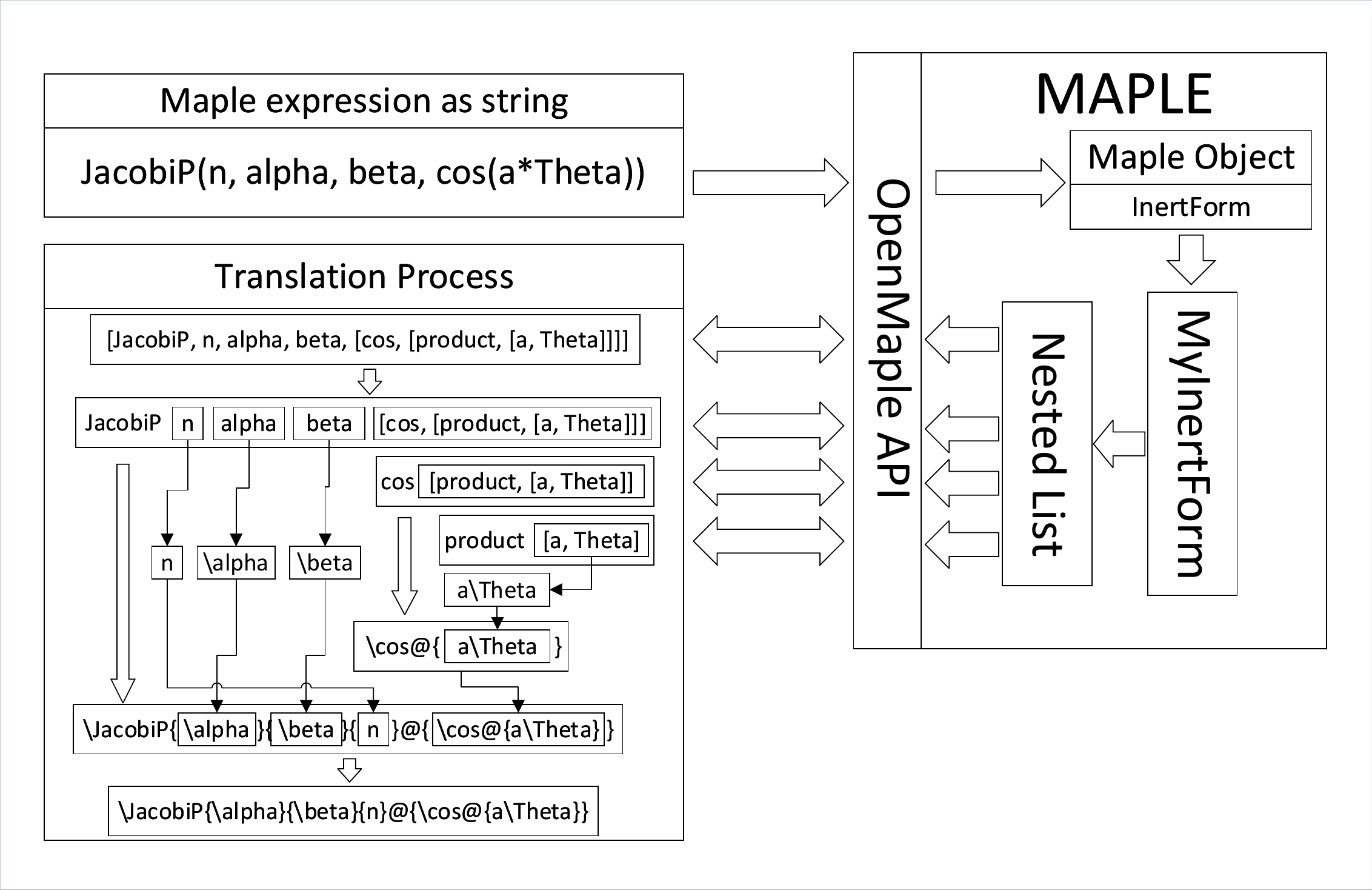}
\caption{The program flow diagram explains the translation 
from \Maple{} to semantic \LaTeX. The input string is parsed into a \Maple{} 
object and \Maple{} procedures create a new internal form of the object and 
builds a {\tt nested list} from this new form. The CAS translation process assembles
the semantic \LaTeX{} expression by translating each element recursively.}
\label{multiplescanapproach2}
\end{figure}

Our CAS translation engine enters the {\tt 1D Maple} representation via
the {\sf OpenMaple} API for {\sf Java} \cite{OpenMapleAPI} and converts
the previously described {\verb|Inert_Form|} to a {\tt nested list} representation.
For \Maple{} expressions, the {\tt nested list} has a tree structure.
We have organized the backward translation in a similar fashion 
to the forward translation (see \S \ref{sec:LatexToCAS}).

Since \Maple{} automatically tries to simplify input expressions, we 
implemented some additional changes to prevent such simplifications 
and changes to the input expression. We would prefer that the representation 
of a translated expression remain as similar as possible to the input expression.
This facilitates user comprehension, as well as the debugging process, of the 
CAS translation.  \Maple's internal representation presents obstacles when 
trying to keep an internal expression in the syntactical form of the input 
expression.  For instance, \Maple{} performs automatic (1) simplification 
of input expressions; (2) representation of radicals as powers with fractional 
exponents (e.g., \verb|\sqrt[5]{x^3}| represented as \verb|x^{3/5}|); 
(3) representation of negative terms as positive terms multiplied by \verb|-1|
(since \Maple{}'s internal structure has no primitives for negation or 
subtraction); and (4) representation of division by a term as a multiplication 
of that term raised to a negative power (since \Maple{}'s internal structure 
has no primitives for division).

To prevent automatic simplifications in \Maple{}, one can enclose input expressions 
between single quotes \textquotesingle ...\textquotesingle,
also known as {\tt unevaluation quotes}. This does not 
prevent arithmetic simplifications but does prevent all other simplifications to the input expression. 
For instance, if we have input \verb|sin(Pi)+2-1|, then the output is \verb|1|; and if we 
have input \textquotesingle\verb|sin(Pi)+2-1|\textquotesingle, then the output is \verb|sin(Pi)+1|.
By using unevaluation quotes, \Maple{} does not convert a radical to a power with fractional 
exponents, and the internal representation remains an unevaluated \verb|sqrt| (for square roots) 
or \verb|root| (for higher order radicals).
\Maple{} automatically represents a negative term such as \verb|-a| by a 
product \verb|a*(-1)|. To resolve this we 
first switch the order of the terms so that constants are in front, e.g., \verb|(-1)*a|, and 
then check if the leading constant is positive or negative. If it is negative, we remove the
multiplication and insert a negative sign in front of the term.

\Maple's rendering engine only changes negative powers to fractions if the 
power is a ratio of integers, otherwise it keeps the exponent representation. 
We only translate terms with negative integer exponents to fractions, 
and otherwise retain the internal exponent representation.  For this purpose, we 
perform a preprocessing step (in \Maple{}) that introduces a new {\tt DIVIDE} element in the 
tree representation.
For instance, without the {\tt DIVIDE} element the input
\verb|(1/(x+3))^(-I)| produces \verb|\left((3+x)^{-1}\right)^{-\iunit}|,
and with the {\tt DIVIDE} element it produces \verb|\left(\frac{1}{3+x}\right)^{-\iunit}|.

\setcounter{footnote}{0}
\setcounter{mpfootnote}{0}
Using the above described manipulations, a typical translated expression is very similar 
to the input expression. As an example, without any of the techniques above, the 
input expression \verb|cos(Pi*2)/sqrt((3*beta)/4-3*I)| would be automatically 
simplified and changed internally, and the resulting semantic \LaTeX\ would be
\verb|2\idt(3\idt\beta+12\idt\iunit\idt(-1))^{-\frac{1}{2}}|.\footnote{\textbackslash{\tt idt} is 
our semantic \LaTeX\ macro which represents multiplication 
without any corresponding presentation appearance.}
With unevaluation quotes, the CAS translation produces \\[0.1cm]
\noindent{\footnotesize\phantom{|\!\!\!\!}\verb|\cos@{\cpi\idt2}\idt\left(\sqrt{\beta\idt\frac34+\iunit\idt(-3)}\right)^{-1}|}.\\
Furthermore, with our improvements for subtractions, we translate the radicand 
to\\ \verb|\frac{3}{4}\idt\beta-3\idt\iunit|, and with the {\tt DIVIDE} element, we translate the 
base with exponent \verb|-1| as a fraction, and our translated expression is\\
\noindent{\footnotesize \phantom{|}\verb|\frac{\cos@{2\idt\cpi}}{\sqrt{\frac{3}{4}\idt\beta-3\idt\iunit}}|},\\
which is very similar to the input expression.

\section{Evaluation}
\label{evaluation}
Here, we describe our approach for validating the correctness of our mappings, as well as discuss 
the performance of our system obtained on a hand crafted test set.

One validation approach is to take advantage of numerical evaluation using software tools such as
the {\sf DLMF Standard Reference Tables} (DLMF Tables) \cite{DLMF:Tables}, CAS, and software 
libraries\footnote{See for instance:~\href{http://dlmf.nist.gov/software/}{http://dlmf.nist.gov/software}.}.
These tools provide numerical evaluation for special functions with their own unique features.  One 
can validate forward CAS translations by comparing numerical values in CAS to ground truth values.

Another validation approach is to use mathematical relations between different
functions.  For instance, if we forward translate two functions separately,
one could determine if the relation between the two translated functions remains valid.
One example relation is for the Jacobi elliptic 
functions ${\rm sn}$, ${\rm cn}$, ${\rm dn}$, and the complete elliptic integral $K$
\cite[Table 22.4.3]{NIST:DLMF}, namely
${\rm sn}(z+K(k),k)={\rm cn}(z,k)/{\rm dn}(z,k),$ where $z\in{\mathbb C},$ and $k\in(0,1)$.
In the limit as $k\to0$, this relation produces 
$\sin@{z+\frac{\pi}{2}} = \cos(z),$
where $z\in{\mathbb C}$.
The DLMF provides relations such as these for many special functions. 
An alternative relation is particularly helpful to validate CAS translations with different positions 
of branch cuts, namely the relation between the parabolic cylinder function $U$ and the modified 
Bessel function of the second kind
\cite[(12.7.10)]{NIST:DLMF}
$U(0,z) = \sqrt{z/(2\pi)}K_{1/4}(\tfrac{1}{4}z^2),$
where $z\in{\mathbb C}$.
Note that $z^2$ is no longer on the principal branch of the modified Bessel function of the second 
kind when 
${\mathrm{ph}}(z) \in (\frac{\pi}{2},\pi)$, 
but a CAS would still compute values on the principal branch. 
Therefore, a CAS translation from \verb|\BesselK{\frac{1}{4}}@{\frac{1}{4}z^2}| to \verb|BesselK(1/4,(1/4)*z^2)|
is incorrect if 
${\mathrm{ph}}(z) \in (\frac{\pi}{2},\pi)$, even though the equation is true in that domain.
In order for the CAS to verify the formula in that domain, it must use \cite[(10.34.4)]{NIST:DLMF} 
for the function on the right-hand side.
Other validation tests may not be able to identify a problem with this CAS translation.

One obstacle for such relations are the limitations of ever-improving 
CAS simplification functions. Define the {formula difference}, as the difference between
the left- and right-hand sides of a mathematical formula. CAS simplify for the Jacobi elliptic/trigonometric 
relation should produce \verb|0|, but might have more difficulties with the parabolic cylinder 
function relation.
However, CAS simplify functions work more effectively on round trip tests.
\subsection{Round trip tests}\label{roundtriptests}
One of the main techniques we use to validate CAS translations are round trip 
tests which take advantage of CAS simplification functions. Since we have developed 
CAS translations between semantic \LaTeX{} $\leftrightarrow$ \Maple, round trip tests are
evaluated in \Maple. 
\Maple's simplification function is called \verb|simplify|. 
Two expressions are symbolically equivalent, if \verb|simplify| returns 
zero for the {formula difference}. On the other hand, it is 
not possible to disprove the equivalence of the expressions when the function 
returns something different to zero.

Our round trip tests start either from a valid semantic \LaTeX{} expression
or from a valid \Maple{} expression. A CAS translation from the start representation
to the other representation and back again is called one cycle. A round trip reaches
a fixed point, when the string representation is identical to its
previous string representation. The round trip test concludes when it reaches a 
fixed point in both representations.  Additionally, we test if the fixed point 
representation in \Maple{} is symbolically equivalent to the input representation by 
simplifying the differences between both of these with the \Maple{} simplify function.
Since there is no mathematical equivalence tester for \LaTeX{} expressions (neither generic 
nor semantic \LaTeX), we manually verify \LaTeX\ representations for our test cases by 
rendering the \LaTeX{}.

\begin{wraptable}[8]{l}[0.0cm]{7.0cm}
\vspace{-0.7cm}
\caption{A round trip test reach a fixed point.}
\begin{tabular}{|c|c|}
\hline step & semantic \LaTeX{}/\Maple{} representations
\\\hline
0 & \verb|\frac{\cos@{a\Theta}}{2}|\TT\\
1 & \verb|(cos(a*Theta))/(2)| \\
2 & \verb|\frac{1}{2}\idt\cos@{a\idt\Theta}| \\
3 & \verb|(1)/(2)*cos(a*Theta)| \\
\hline
\end{tabular}
\label{fixedpointtablehere}
\end{wraptable}

As shown in \S\ref{sec:backwardTranslationImplementation}, prior to backward 
translation, in round trip testing, there will be differences between input 
and output \Maple{} representations.  After adapting these changes, 
and assuming the functions exist in both semantic \LaTeX\ and CAS, the round 
trip test should reach a fixed point. In fact, we reached a fixed point in 
semantic \LaTeX{} after one cycle and in \Maple{} after $1\tfrac12$ cycles 
(see Table \ref{fixedpointtablehere} for an example) for most of the cases we tried.
If the input representation is 
already identical to \Maple's representation, then the fixed point will be reached 
after at most a half cycle.

One example exception is for CAS translations which introduce additional function 
compositions on arguments.  For instance, Legendre's incomplete elliptic 
integrals \cite[(19.2.4-7)]{NIST:DLMF} are defined with the 
amplitude $\phi$ in the first argument, while \Maple's implementation takes the trigonometric
sine of the amplitude as the first argument.  For instance, one has the CAS translations
\verb|\EllIntF@{\phi}{k}| $\mapsto$ \verb|EllipticF(sin(phi),k)|, and 
\verb|\EllIntF@{\asin@{z}}{k}| $\mapsfrom$ \verb|EllipticF(z,k)|.
These CAS translations produce an infinite chain of sine and inverse sine function calls.
Because round trip tests prevent simplification during the 
translation process (see \S\ref{sec:backwardTranslationImplementation}), \Maple{} is not 
used to simplify the chain until the round trip test is concluded.
\subsection{Summary of evaluation techniques}
\label{Summaryofevaluationtechniques}
Equivalence tests for special function relations are able to verify relations
in CAS as well as identify hidden problems such as differences in branch cuts and CAS 
limitations.  We use the simplify method to test equivalences. 
For the relations in \S\ref{evaluation}, CAS simplify for the Jacobi elliptic function 
example yields \verb|0|.  Furthermore, a spectrum of real, complex, and complex conjugate 
numerical values for $z$ and $k\in(0,1)$ the {formula difference} converges to zero 
for an increasing precision.  If simplification returns something other than zero, we can 
test the equivalence for specific values.  For the Bessel function relation, the 
{formula difference} for $z=1+i$ converges to zero for increasing precision, but 
does not converge to zero if $z=-1+i$. However, using analytic continuation 
\cite[(10.34.4)]{NIST:DLMF}, it does converges to zero. Clearly, the numerical evaluation 
test is also able to locate branch cut issues in the CAS translation.  Furthermore, this 
provides a very powerful debugging method for our translation as well as for CAS functionality.
This was demonstrated by discovering an overall sign error in DLMF equation \cite[(14.5.14)]{NIST:DLMF}.

Round trip tests are also useful for identifying 
syntax errors in the semantic \LaTeX\ since the CAS translation then fails. The simplification 
procedure is improved for round trip tests, because it only needs to simplify similar expressions with 
identical function calls. However, this approach is not able to identify hidden problems that 
a CAS translation might need to resolve in order to be correct, if the round trip test has 
not reached a fixed point.  Other than with the round trip test approach, we have not 
discovered any automated tests for backward CAS translations.
We have evaluated 37 round trip test cases which produce a fixed point, 
similar to that given in Table \ref{fixedpointtablehere}.  
These use formulae from the DLMF/DRMF and produce a difference of the left- and right-hand 
sides equaling \verb|0|. 

We have created a test dataset\footnote{We are planning to make the dataset
available from \href{http://drmf.wmflabs.org/}{http://drmf.wmflabs.org}.} of 4,165 semantic \LaTeX\ formulae, extracted from 
the DLMF.  We translated each test case to a representation in \Maple{} and used \Maple's 
{\tt simplify} function on the formula difference to verify that the 
translated formulae remain valid. Our forward translation tool (\S\ref{sec:LatexToCAS})
was able to translate 2,232 (approx. 53.59\%) test cases and verify 477 of these.
Pre-conversion improved the effectiveness of {\tt simplify} and were used to 
convert the translated expression to a different form before simplification of the formula 
difference. We used conversions to exponential and hypergeometric form and expanded the 
translated expression. Pre-conversion increased the number of formulae verified to 662 and 
1,570 test cases were translated but not verified. The remaining 1,933 test cases were not 
translated, because they contain DLMF/DRMF macros without a known translation to \Maple{} 
(987 cases), such as the $q$-hypergeometric function \cite[(17.4.1)]{NIST:DLMF} (in 58 cases), 
or an error appeared during the translation or verification process (639 cases). Furthermore, 
316 cases were ignored, because they did not contain enough semantic information to provide a 
translation or the test case was not a relation. It is interesting to note that we were able to
enhance the semantics of 74 Wronskian relations by rewriting the macro so that it included the variable 
that derivatives are taken with respect to as a parameter. A similar semantic enhancement 
is possible for another 186 formulae where
the potentially ambiguous prime 
notation `\hspace{0.03cm}\textquotesingle\hspace{0.03cm}' is used for derivatives.
\\[0.1cm]
\noindent 
\hspace{-0.111cm}
{\bf\large\!\hspace{0.04cm}Acknowledgements\hspace{0.02cm}}\footnote{The mention of specific products, trademarks, or brand
names is for purposes of identification only. Such mention is not to be interpreted in any way
as an endorsement or certification of such products or brands by the National Institute of
Standards and Technology, nor does it imply that the products so identified are necessarily
the best available for the purpose. All trademarks mentioned herein belong to their
respective owners.}
We are indebted to Wikimedia Labs, the XSEDE project,\\[0.25cm] Springer-Verlag,
the California Institute of Technology, and  Maplesoft for their
contributions and continued support.  We would also like to thank
Eric Weisstein for supplying the Wolfram {\tt eCF} dataset,
Annie Cuyt, Franky Backeljauw, and Stefan Becuwe for supplying the 
University of Antwerp {\tt CFSF} {\sf Maple} dataset, and Adri Olde Daalhuis for discussions
related to complex multivalued functions.

\label{sect:bib}
\printbibliography[keyword=primary]

\end{document}